\documentclass[aps,prl,showpacs,twocolumn,groupedaddress]{revtex4}  
\usepackage{graphicx}  
\usepackage{dcolumn}   
\usepackage{bm}        
\usepackage{amssymb}   
\usepackage{epsfig}

\newcommand{\ipb}{\ensuremath{\mathrm{pb^{-1}}}}

\newcommand{\ttb}{\ensuremath{t\bar{t}}}
\newcommand{\bb}{\ensuremath{b\bar{b}}}
\newcommand{\cc}{\ensuremath{c\bar{c}}}
\newcommand{\pp}{\ensuremath{p\bar{p}}}
\newcommand{\W}{{\it W}}
\newcommand{\met}{\ensuremath{\rlap{\kern0.25em/}E_T}}
\newcommand{\wev}{\ensuremath{W\rightarrow e\nu}}
\newcommand{\wmv}{\ensuremath{W\rightarrow \mu\nu}}

\newcommand{\zmm}{\ensuremath{Z\rightarrow \mu\mu}}
\newcommand{\ejet}{\ensuremath{e+\rm{jets}}}
\newcommand{\mjet}{\ensuremath{\mu + \rm{jets}}}
\newcommand{\dr}{\ensuremath{\mathcal{R}}}

\def\lsim{\mathrel{\rlap{\lower4pt\hbox{\hskip1pt$\sim$}}
    \raise1pt\hbox{$<$}}}                
\def\gsim{\mathrel{\rlap{\lower4pt\hbox{\hskip1pt$\sim$}}
    \raise1pt\hbox{$>$}}}                

\begin{document}



\title{\boldmath A Search for Anomalous Heavy-Flavor Quark Production\\in Association with $W$ Bosons}
%
\author{                                                                      
V.M.~Abazov,$^{34}$                                                           
B.~Abbott,$^{71}$                                                             
M.~Abolins,$^{62}$                                                            
B.S.~Acharya,$^{28}$                                                          
M.~Adams,$^{49}$                                                              
T.~Adams,$^{47}$                                                              
M.~Agelou,$^{17}$                                                             
J.-L.~Agram,$^{18}$                                                           
S.H.~Ahn,$^{30}$                                                              
M.~Ahsan,$^{56}$
% ...
\\                                                                            
\vskip 0.30cm                                                                 
\centerline{(D\O\ Collaboration)}                                             
\vskip 0.30cm                                                                 
}                                                                             
\address{                                                                     
\centerline{$^{1}$Universidad de Buenos Aires, Buenos Aires, Argentina}       
\centerline{$^{2}$LAFEX, Centro Brasileiro de Pesquisas F{\'\i}sicas,         
                  Rio de Janeiro, Brazil}                                     
\centerline{$^{3}$Universidade do Estado do Rio de Janeiro,                   
                  Rio de Janeiro, Brazil}                                     
\centerline{$^{4}$Instituto de F\'{\i}sica Te\'orica, Universidade            
                  Estadual Paulista, S\~ao Paulo, Brazil}                     
\centerline{$^{5}$University of Alberta, Edmonton, Alberta, Canada,           
               Simon Fraser University, Burnaby, British Columbia, Canada,}   
\centerline{York University, Toronto, Ontario, Canada, and                    
         McGill University, Montreal, Quebec, Canada}                         
\centerline{$^{6}$Institute of High Energy Physics, Beijing,                  
                  People's Republic of China}                                 
\centerline{$^{7}$Universidad de los Andes, Bogot\'{a}, Colombia}             
\centerline{$^{8}$Center for Particle Physics, Charles University,            
                  Prague, Czech Republic}                                     
\centerline{$^{9}$Czech Technical University, Prague, Czech Republic}         
\centerline{$^{10}$Institute of Physics, Academy of Sciences, Center          
                  for Particle Physics, Prague, Czech Republic}               
\centerline{$^{11}$Universidad San Francisco de Quito, Quito, Ecuador}        
\centerline{$^{12}$Laboratoire de Physique Corpusculaire, IN2P3-CNRS,         
                 Universit\'e Blaise Pascal, Clermont-Ferrand, France}        
\centerline{$^{13}$Laboratoire de Physique Subatomique et de Cosmologie,      
                  IN2P3-CNRS, Universite de Grenoble 1, Grenoble, France}     
\centerline{$^{14}$CPPM, IN2P3-CNRS, Universit\'e de la M\'editerran\'ee,     
                  Marseille, France}                                          
\centerline{$^{15}$Laboratoire de l'Acc\'el\'erateur Lin\'eaire,              
                  IN2P3-CNRS, Orsay, France}                                  
\centerline{$^{16}$LPNHE, IN2P3-CNRS, Universit\'es Paris VI and VII,         
                  Paris, France}                                              
\centerline{$^{17}$DAPNIA/Service de Physique des Particules, CEA, Saclay,    
                  France}                                                     
\centerline{$^{18}$IReS, IN2P3-CNRS, Universit\'e Louis Pasteur, Strasbourg,  
                France, and Universit\'e de Haute Alsace, Mulhouse, France}   
\centerline{$^{19}$Institut de Physique Nucl\'eaire de Lyon, IN2P3-CNRS,      
                   Universit\'e Claude Bernard, Villeurbanne, France}         
\centerline{$^{20}$III. Physikalisches Institut A, RWTH Aachen,               
                   Aachen, Germany}                                           
\centerline{$^{21}$Physikalisches Institut, Universit{\"a}t Bonn,             
                  Bonn, Germany}                                              
\centerline{$^{22}$Physikalisches Institut, Universit{\"a}t Freiburg,         
                  Freiburg, Germany}                                          
\centerline{$^{23}$Institut f{\"u}r Physik, Universit{\"a}t Mainz,            
                  Mainz, Germany}                                             
\centerline{$^{24}$Ludwig-Maximilians-Universit{\"a}t M{\"u}nchen,            
                   M{\"u}nchen, Germany}                                      
\centerline{$^{25}$Fachbereich Physik, University of Wuppertal,               
                   Wuppertal, Germany}                                        
\centerline{$^{26}$Panjab University, Chandigarh, India}                      
\centerline{$^{27}$Delhi University, Delhi, India}                            
\centerline{$^{28}$Tata Institute of Fundamental Research, Mumbai, India}     
\centerline{$^{29}$University College Dublin, Dublin, Ireland}                
\centerline{$^{30}$Korea Detector Laboratory, Korea University,               
                   Seoul, Korea}                                              
\centerline{$^{31}$CINVESTAV, Mexico City, Mexico}                            
\centerline{$^{32}$FOM-Institute NIKHEF and University of                     
                  Amsterdam/NIKHEF, Amsterdam, The Netherlands}               
\centerline{$^{33}$University of Nijmegen/NIKHEF, Nijmegen, The               
                  Netherlands}                                                
\centerline{$^{34}$Joint Institute for Nuclear Research, Dubna, Russia}       
\centerline{$^{35}$Institute for Theoretical and Experimental Physics,        
                  Moscow, Russia}                                             
\centerline{$^{36}$Moscow State University, Moscow, Russia}                   
\centerline{$^{37}$Institute for High Energy Physics, Protvino, Russia}       
\centerline{$^{38}$Petersburg Nuclear Physics Institute,                      
                   St. Petersburg, Russia}                                    
\centerline{$^{39}$Lund University, Lund, Sweden, Royal Institute of          
                   Technology and Stockholm University, Stockholm,            
                   Sweden, and}                                               
\centerline{Uppsala University, Uppsala, Sweden}                              
\centerline{$^{40}$Lancaster University, Lancaster, United Kingdom}           
\centerline{$^{41}$Imperial College, London, United Kingdom}                  
\centerline{$^{42}$University of Manchester, Manchester, United Kingdom}      
\centerline{$^{43}$University of Arizona, Tucson, Arizona 85721, USA}         
\centerline{$^{44}$Lawrence Berkeley National Laboratory and University of    
                  California, Berkeley, California 94720, USA}                
\centerline{$^{45}$California State University, Fresno, California 93740, USA}
\centerline{$^{46}$University of California, Riverside, California 92521, USA}
\centerline{$^{47}$Florida State University, Tallahassee, Florida 32306, USA} 
\centerline{$^{48}$Fermi National Accelerator Laboratory, Batavia,            
                   Illinois 60510, USA}                                       
\centerline{$^{49}$University of Illinois at Chicago, Chicago,                
                   Illinois 60607, USA}                                       
\centerline{$^{50}$Northern Illinois University, DeKalb, Illinois 60115, USA} 
\centerline{$^{51}$Northwestern University, Evanston, Illinois 60208, USA}    
\centerline{$^{52}$Indiana University, Bloomington, Indiana 47405, USA}       
\centerline{$^{53}$University of Notre Dame, Notre Dame, Indiana 46556, USA}  
\centerline{$^{54}$Iowa State University, Ames, Iowa 50011, USA}              
\centerline{$^{55}$University of Kansas, Lawrence, Kansas 66045, USA}         
\centerline{$^{56}$Kansas State University, Manhattan, Kansas 66506, USA}     
\centerline{$^{57}$Louisiana Tech University, Ruston, Louisiana 71272, USA}   
\centerline{$^{58}$University of Maryland, College Park, Maryland 20742, USA} 
\centerline{$^{59}$Boston University, Boston, Massachusetts 02215, USA}       
\centerline{$^{60}$Northeastern University, Boston, Massachusetts 02115, USA} 
\centerline{$^{61}$University of Michigan, Ann Arbor, Michigan 48109, USA}    
\centerline{$^{62}$Michigan State University, East Lansing, Michigan 48824,   
                   USA}                                                       
\centerline{$^{63}$University of Mississippi, University, Mississippi 38677,  
                   USA}                                                       
\centerline{$^{64}$University of Nebraska, Lincoln, Nebraska 68588, USA}      
\centerline{$^{65}$Princeton University, Princeton, New Jersey 08544, USA}    
\centerline{$^{66}$Columbia University, New York, New York 10027, USA}        
\centerline{$^{67}$University of Rochester, Rochester, New York 14627, USA}   
\centerline{$^{68}$State University of New York, Stony Brook,                 
                   New York 11794, USA}                                       
\centerline{$^{69}$Brookhaven National Laboratory, Upton, New York 11973, USA}
\centerline{$^{70}$Langston University, Langston, Oklahoma 73050, USA}        
\centerline{$^{71}$University of Oklahoma, Norman, Oklahoma 73019, USA}       
\centerline{$^{72}$Brown University, Providence, Rhode Island 02912, USA}     
\centerline{$^{73}$University of Texas, Arlington, Texas 76019, USA}          
\centerline{$^{74}$Southern Methodist University, Dallas, Texas 75275, USA}   
\centerline{$^{75}$Rice University, Houston, Texas 77005, USA}                
\centerline{$^{76}$University of Virginia, Charlottesville, Virginia 22901,   
                   USA}                                                       
\centerline{$^{77}$University of Washington, Seattle, Washington 98195, USA}  
}                                                                             
\date{\today}

\begin{abstract}

We present a search for anomalous production of heavy-flavor quark
jets in association with a $W$ boson at the Fermilab Tevatron $\pp$
Collider. This search is conducted through an examination of the
exclusive jet spectrum of $W+$jets final states in which the
heavy-flavor quark content has been enhanced by requiring at least one
tagged jet in an event. Jets are tagged by the combined use of two
algorithms, one based on semileptonic decays of $b/c$ hadrons, and the
other on their lifetimes. We compare data in $\ejet$ (164~$\ipb$) and
$\mjet$ (145~$\ipb$) channels, collected with the D\O\ detector at
$\sqrt{s}=1.96$~TeV, to expectations from the standard model, and set
upper limits on anomalous production of such events.
\end{abstract}
 \maketitle

The heavy-flavor (HF) content of jets produced in association with a
$W$ boson in $\pp$ collisions provides a test of the standard model
(SM), and an excess would suggest a non-SM source of physics. The CDF
collaboration recently reported just such an excess in the exclusive
$W+$ HF-jet spectrum in which one jet was tagged using both
secondary-vertex (SVT) and soft-lepton (SLT) tagging
algorithms~\cite{CDFanom2}. To check for the presence of this anomaly
in our data, we also select jets tagged with both algorithms. In
addition, we use two benchmark SM processes as models for new physics
and derive upper limits on such processes.

At the Tevatron, the primary SM contributions to a $W$ boson
associated with HF quarks in the final state are expected to be from
$\ttb$, $W\bb/\cc$ (where the $\bb$ or $\cc$ pairs arise from gluon
splitting), and $Wc$ final states, with additional contributions
arising from single top quark or $WZ$ $(\textrm{with }Z\rightarrow
\bb/\cc)$ production. The production of $W$ bosons accompanied by
light quarks or gluons (referred to as $W+$jets in this Letter)
contributes to the background when the light-quark or gluon jets are
misidentified as jets from HF quarks.  Since $W$ bosons are identified
through their $\wev$ and $\wmv$ decays, background can arise from
$Z\bb$, $ZZ$ $(\textrm{with one }Z\rightarrow \bb/\cc)$, and $Z+$jets
production when one of the leptons from the $Z\rightarrow
\ell^{+}\ell^{-}$ decay is not observed in the detector. The main
instrumental background arises from multijet processes in which a jet
is misidentified as a lepton, and an imbalance in transverse momentum
($\met$) is generated through a mismeasurement of the jets or a
lepton.  To be selected, these kinds of events must also contain
tagged HF jets or misidentified non-HF jets.

The data were collected with the D\O\ detector~\cite{run2det} during
Run II of the Fermilab Tevatron Collider in $\pp$ collisions at
$\sqrt{s}=1.96$~TeV. The components used in this analysis include the
central tracker, calorimeter, and muon detectors. The central tracker
consists of a silicon microstrip tracker (SMT) and a central fiber
tracker (CFT), both located within a 2~T superconducting solenoidal
magnet. The uranium/liquid-argon calorimeter consists of three
sections, each housed in a separate cryostat~\cite{run1det}. The
central calorimeter (CC) covers pseudorapidity $|\eta| \lsim 1.1$,
while the two end calorimeters (EC) extend the coverage to $|\eta|
\approx 4.0$. The muon system is located outside the calorimeters, and
consists of a layer of tracking detectors and scintillation trigger
counters inside 1.8~T iron toroids, followed by two more similar
layers outside the toroids.

The $\wev$ and $\wmv$ decay candidates are selected initially by
triggering on electrons and muons. The average trigger efficiency for
electrons with transverse momentum $p_{T} >20\textrm{ GeV}/c$ and
$\left| \eta\right| < 1.1$ is (97.0$\pm$0.3)\%. The average trigger
efficiency for muons with $p_{T}>20\textrm{ GeV}/c$ and $\left| \eta
\right| < 1.6$ is (62.1$\pm$3.4)\%. The integrated luminosity is $164
\pm 11~\ipb$ for the electron sample and $145 \pm 9~\ipb$ for the muon
sample.

Candidate events for $\wev$ decays are selected by requiring exactly
one isolated electron with $p_{T}>20\textrm{ GeV}/c$ and $\left| \eta
\right| < 1.1$, defined relative to the geometrical center of the
detector. Lepton isolation requires a separation in $\eta$ and azimuth
($\phi$) of $\dr = \sqrt{(\Delta\eta)^{2} + (\Delta\phi)^{2}}>0.5$
from all jets in the event. Electrons are defined using a cone
algorithm, and by the energies deposited in calorimeter towers within
a radius of $\dr = 0.2$ of the electron axis, with at least 90\%
required to be within the electromagnetic portion of the calorimeter,
and by the total energy in a cone of $\dr=0.4$ centered on the same
axis, which must not exceed by more than 15\% the reconstructed
electron's energy. In addition, the longitudinal and transverse shower shape
must be compatible with that expected from an electron.

Candidate events for $\wmv$ decays must contain exactly one isolated
muon with $p_{T} >20\textrm{ GeV}/c$ and $\left| \eta \right| < 1.6$,
also defined relative to the geometrical center of the detector. Muons
are required to satisfy two additional isolation criteria: the
transverse energy deposited in the calorimeter in the annular region
of $0.1<\dr <0.4$ around the muon's path must be smaller than 2.5~GeV;
and the vector sum of the $p_T$ values of all tracks within $\dr=0.5$
of the muon's trajectory must be less than $2.5\textrm{ GeV}/c$
(excluding the track matched to the muon's trajectory).

Lepton identification is refined by requiring the trajectory of a
track reconstructed in the SMT and CFT to match either the position of
the electron energy cluster in the calorimeter or the position of hits
in the muon detector. To complete the selection, all events are also
required to have $\met >20\textrm{ GeV}$, and the azimuthal angle
between the lepton and the direction of the $\met$ must be greater
than $\pi / 8$. To eliminate poorly reconstructed events, the primary
vertex (PV) of the event must contain at least three tracks, and its
$z$-position (along the beam) has to be closer than 60~cm from the
center of the detector. Finally, to reject multijet background, we
require a reconstructed transverse mass consistent with that of the
$W$ boson, $40<M_{W_{T}}<120 \textrm{ GeV}/c^{2}$. In calculating
$M_{W_{T}}$, we assume that the $\met$ corresponds to the transverse
energy of the neutrino.

Upon selection of $W$-boson candidates, we evaluate the HF-quark
content of each event. Jets are defined using an iterative seed-based
cone algorithm (including mid-points), clustering calorimeter energy
within $\dr=0.5$. This is subsequently corrected for jet energy scale,
based on momentum balance in photon+jet events~\cite{jetalg}. We
consider only jets with $E_{T}>25\textrm{ GeV}/c$ and $\left| \eta
\right| < 2.5$. These jets are then evaluated using two HF-tagging
algorithms, as described below.

The soft-lepton tagging (SLT) algorithm is based on low-$p_{T}$ muons
arising from semileptonic decays of HF quarks (via virtual $W$ bosons)
that are produced near a jet in $(\eta,\phi)$ space. Only muons with
$p_{T}>4\textrm{ GeV}/c$ and $\left| \eta \right| < 2.0$ are
considered. To reject $\zmm$ background, we require $p_{T} <15\textrm{
GeV}/c$ for the muon. Jets with a muon within $\dr=0.5$ of the jet
axis are deemed tagged. Typical SLT efficiencies for $b$-quark jets
are approximately 11\%, and 0.4\% for light-quark jets. The additional
muon present in SLT events causes an increase in the average
single-muon trigger efficiency from ($62.1\pm3.4$)$\%$ to
($68.4\pm3.5$)$\%$.

Secondary-vertex tagging (SVT) is used to identify displaced decay
vertices of long-lived particles. To form secondary vertices (SV),
charged tracks are selected on the basis of the significance of their
distance-of-closest-approach (dca) to the PV. Tracks are first grouped
in $\mathcal{R}=0.5$ cones around a seed track with $p_{T}>1$ GeV$/c$
and $\rm{dca}/\sigma_{\rm{dca}}>3.5$, where $\sigma_{\rm{dca}}$ is the
uncertainty on the track's dca. Proto-vertices are formed by adding
tracks to the initial grouping, provided their contribution to the
$\chi^{2}$ of the vertex fit is small. Secondary vertices are selected
by requiring the transverse distance from the SV to the beam
direction, $L_{xy}$, to be less than 2.6~cm, and the decay-length
significance, $\frac{L_{xy}}{\sigma_{L_{xy}}}$ to be greater than 7,
where $\sigma_{L_{xy}}$ is the estimated uncertainty on $L_{xy}$
calculated from the error matrices of the tracks in the vertex. Jets
are considered tagged by this algorithm when a SV lies within
$\dr=0.5$ of the original jet axis. This SVT algorithm exhibits a
typical tagging rate for $b$-quark jets of $\simeq32\%$, and
0.25\% for light-quark jets.

To predict SM rates, Monte Carlo (MC) events are generated for the
processes mentioned above, with the exception of multijet production,
which is estimated from data as described below. $W/Z$+jets (both HF
and light-quark jets), $\ttb$, and diboson processes are simulated
with $\textsc{alpgen}$~\cite{cite:alpgen}. Single top quark processes
are simulated using $\textsc{comphep}$~\cite{cite:comphep}. All events
are generated with $m_{top} = 175\textrm{ GeV}/c^2$. Hadronization and
showering of these events is based on
\textsc{pythia}~\cite{cite:pythia}. The exceptions are $W/Z+b$
processes, where $Zb$ is simulated using \textsc{pythia}, and the
contribution from $Wb$ is estimated from the parameterized
$\textsc{mcfm}$ MC~\cite{MCFM} and used to calculate a cross section
relative to $W\bb$ production assuming the jet-$p_{T}$ spectrum from
\textsc{pythia} for inclusive $W$ boson production. All MC events are
generated at $\sqrt{s}=1.96\textrm{ TeV}$, using
$\textsc{cteq5l}$~\cite{cite:CTEQ5L} parton distribution functions and
a detailed detector simulation based on
\textsc{geant}~\cite{geant}. To simulate the effect of multiple
interactions in beam crossings, a Poisson-distributed minimum-bias
event overlay, with an average of 0.8 events, is included for all
events.  To avoid an incorrect combination of cross sections among
simulated $W/Z+$jets samples, only events with the same number of
reconstructed jets as the number of initial partons are retained. The
background from multijet events, in which a jet is misidentified as a
lepton, is evaluated using the ``matrix method'' as follows.  Two
samples of $W+$jets event candidates are used: a ``tight'' sample, for
which the lepton identification criteria are as described above, and a
``loose'' sample in which some of these identification criteria are
relaxed.  The probabilities for true leptons to be identified as
loose, and jets as tight leptons (as determined from independent
studies of samples of pure leptons and pure jets) yields the fractions
of true leptons and of misidentified jets in the tight and loose
samples.

\begin{figure}[htb]
{\centering \resizebox*{0.95\columnwidth}{!}{\includegraphics{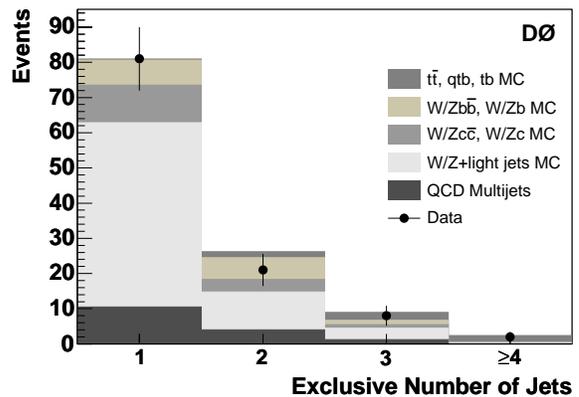}}
\par}
\caption{Exclusive jet multiplicity for $W$-boson candidate events
with at least one SLT-tagged jet. The fourth bin represents the
sum of events containing four or more jets.\label{fig:slt_25} }
\end{figure}

\begin{figure}[htb]
{\centering \resizebox*{0.95\columnwidth}{!}{\includegraphics{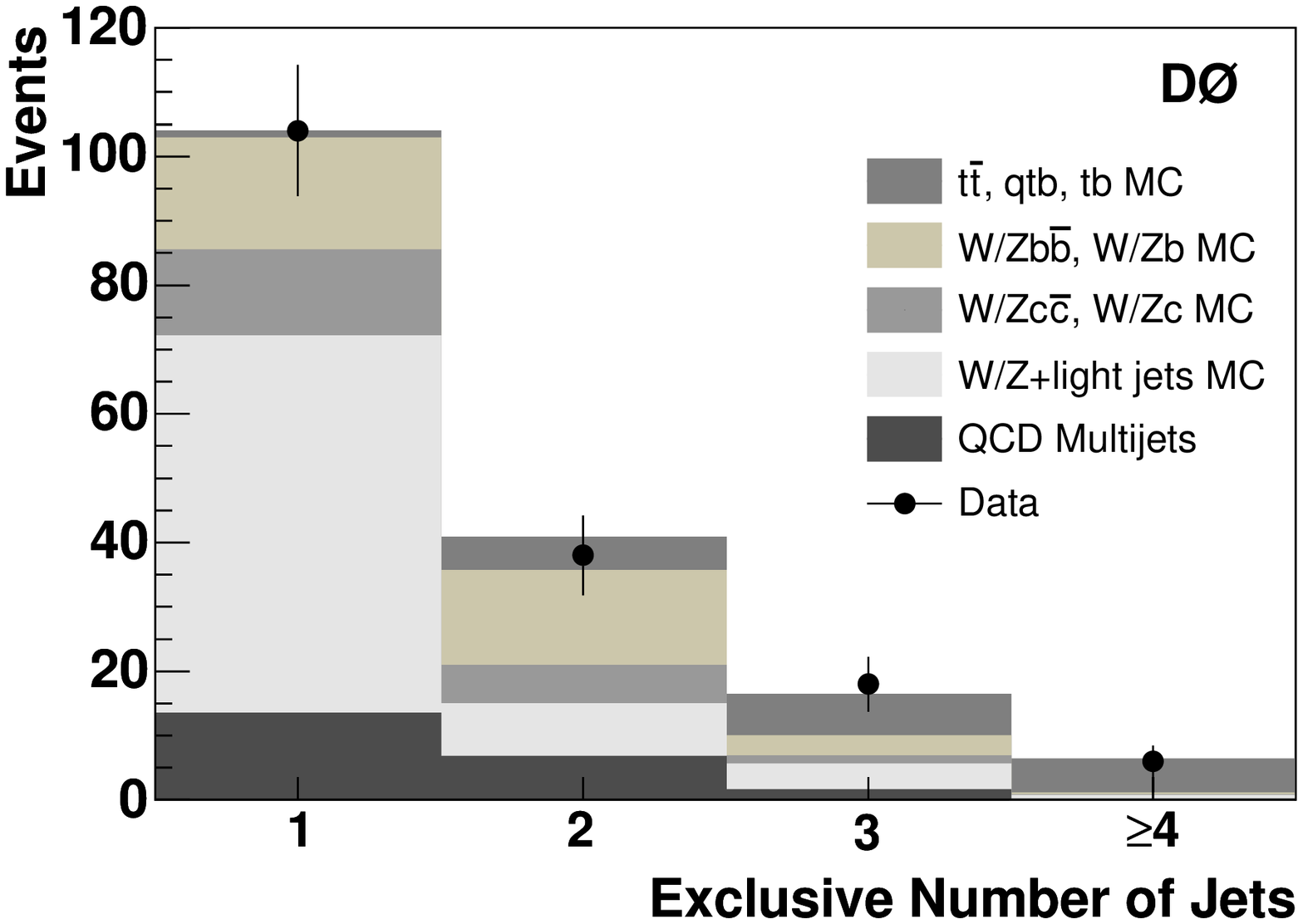}} \par}
\caption{Exclusive jet multiplicity for $W$-boson candidate events
with at least one SVT-tagged jet. The fourth bin represents the
sum of events containing four or more jets.\label{fig:svt_25} }
\end{figure}

\begin{figure}[htb]
{\centering \resizebox*{0.95\columnwidth}{!}{\includegraphics{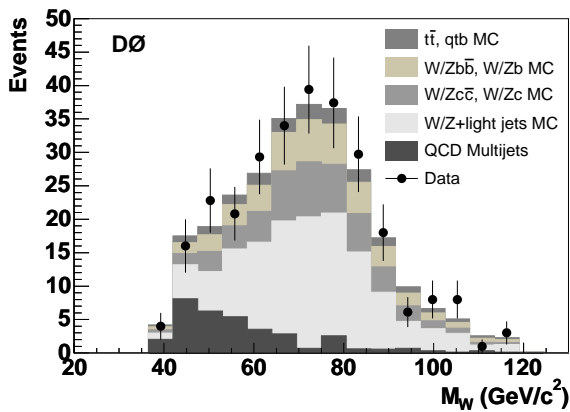}} \par}
\caption{Transverse $W$-boson mass for events containing at
least one SLT- or SVT-tagged jet.\label{fig:tagmass}}
\end{figure}

\begin{figure}[htb]
{\centering \resizebox*{0.95\columnwidth}{!}{\includegraphics{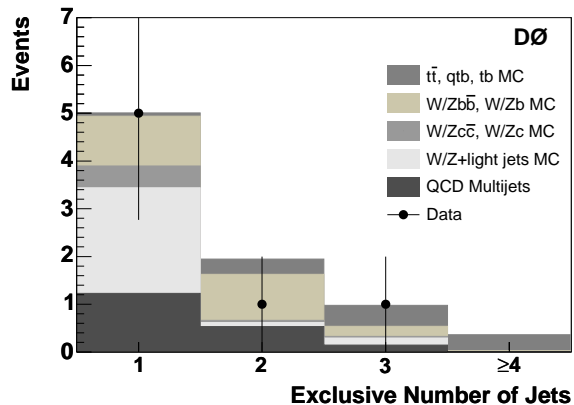}}
\par}
\caption{Exclusive jet multiplicity of $W$-boson candidate events with
at least one jet tagged with both the SVT and SLT algorithms. The
fourth bin represents the sum of events containing four or more
jets.\label{fig:super_25}}
\end{figure}

After the $W$-boson and jet selections, we apply the two HF-tagging
algorithms to the jets. The MC samples are normalized to the
appropriate luminosity, and corrected for differences in HF-tagging
and lepton-identification efficiencies relative to data. Also,
discrepancies in trigger efficiency between MC and data are
corrected for each set of selections. In the following, the $\ejet$
and $\mjet$ samples are combined. Figures~\ref{fig:slt_25}
and~\ref{fig:svt_25} show the exclusive number of jets in events with
at least one SLT-tagged jet and at least one SVT-tagged jet,
respectively. The transverse mass for $W$-boson candidate events
containing at least one SLT or SVT-tagged jet, shown in
Fig.~\ref{fig:tagmass}, agrees well with SM expectation. The
distribution for events with at least one jet tagged with both
algorithms is shown in Fig.~\ref{fig:super_25}.

The dominant sources of experimental uncertainty are common to both
the $\ejet$ and $\mjet$ selections: (\textit{i}) a 6.5\% uncertainty
on the integrated luminosity, (\textit{ii}) a 6\% per jet uncertainty
arising from jet-energy-scale corrections and jet identification,
(\textit{iii}) a 10\% per jet uncertainty arising from the HF-tagging
algorithms, and (\textit{iv}) a 10--18\% uncertainty on the predicted
MC cross sections (depending on sample). The total systematic
uncertainty on the $\ttb$ background is 16\%, 21\% on single top-quark
backgrounds, and 22\% on the $W/Z$+jets backgrounds.

No excess is observed in the ``doubly-tagged'' jet sample. We
therefore proceed to set a limit on the rate anomalous HF-quark
production in association with a $W$ boson. Because we do not propose a
model for such production, we do not base this limit on any
specific efficiency or jet spectrum. We quote limits on the number of
events beyond SM expectation per exclusive jet bin. The 95\%
confidence level (C.L.) upper limits for additional event production
in each bin are shown in Table~\ref{tab:clstab}. These limits are
calculated using a modified Frequentist (CL$_s$) method~\cite{cls}.

\begin{table}[htb]
\begin{center}
\caption{The numbers of observed and predicted $W$-boson events with
at least one jet tagged by both the SLT and SVT algorithms, as a
function of exclusive jet multiplicity.  Also shown are the 95\%
C.L. limits in the form of additional events.\label{tab:clstab}}
\begin{tabular}{lcccc}
\hline \hline
Source  & 1 jet & 2 jets & 3 jets & $\ge$4 jets \\\hline 
Data observation & 5 & 1 & 1 & 0\\ 
SM prediction  & 5.0$\pm$1.2 & 2.0$\pm$0.5 & 1.0$\pm$0.2 & 0.4$\pm$0.06\\
95\% C.L. Limit (events) & 6.7& 3.9& 4.1& 3.0\\\hline \hline
\end{tabular}
\end{center}
\end{table}

Assuming that anomalous HF production has the same event topology as
certain SM process, the above limits can be translated into limits on
cross sections. To this end, we consider two benchmark scenarios:

\begin{enumerate}
\item ``\W\bb-like'' production in which two $b$ quarks are produced
in association with a $W$ boson. In this scenario, additional light
quarks or gluons can be produced, and thereby shift the event topology to more
than two jets. Jets not within the acceptance of the detector can also
cause the event topology to drop to less than two jets. We model this
production using efficiencies for SM $W/Z+\bb$ production.
\item ``Top-like'' production in which a heavy particle is produced
and decays to a \W\ boson and a $b$ quark. An event can
contain two such heavy particles (``\ttb-like'') or one heavy quark
(``single-top-like''), with additional light or heavy quarks and
gluons possible for both cases. We model this scenario using the
cross-section weighted efficiencies for SM \ttb\ and single top-quark
production combined.
\end{enumerate}

We calculate a limit on exclusive jet production for each scenario,
but first ignore the probability for reconstructing the predicted
number of jets, providing a model-independent comparison of
processes with specific jet topologies. The remaining efficiency
represents the effect of $W$-boson selection and HF tagging, and
limits for a specific model can be extracted by multiplying this
value by the efficiency to reconstruct the number of jets found in
each exclusive jet bin. These results are shown in
Table~\ref{tab:xsecNjet}. To evaluate an upper cross-section limit on
inclusive jet production for each scenario, we reintroduce the
efficiency for reconstructing the predicted jets. For inclusive
$\W\bb$-like anomalous production, we sum the first two $W+$jets bins,
as the contribution from the remaining bins is negligible. For
top-like anomalous production, we sum all $W+$jets bins, except the
$n=1$ bin, where the contribution is again
negligible. Table~\ref{tab:limits} shows the 95\% C.L. event limits
for the combinations of jet bins for these two hypotheses, and also
the corresponding anomalous HF production cross-section limits. The
jet reconstruction efficiency is included in the calculations, and the
limits contain the expected efficiencies for the specified SM processes.

\begin{table}[htb]
\begin{center}
\caption{Cross-section upper limits in pb, based on the hypotheses of
``$W \bb$-like'' and ``top-like'' anomalous production of exclusive
number of jets. Each value must still be corrected for the efficiency
of reconstructing the predicted number of jets.\label{tab:xsecNjet}}
\begin{tabular}{lcccc}
\hline \hline 
Model & 1 jet & 2 jets & 3 jets & $\ge$4 jets\\\hline 
$W \bb$-like & 35.0& 9.2& 6.0& 4.5\\ 
Top-like & 12.6& 8.0& 11.3& 15.4\\ \hline \hline
\end{tabular}
\end{center}
\end{table}

\begin{table}[htb]
\begin{center}
\caption{95\% C.L. limits for the number of events summed over the
indicated jet bins. Also shown are cross-section limits based on the
hypotheses of ``$W \bb$-like'' and ``top-like'' anomalous production
for the selected number of jets.\label{tab:limits}}
\begin{tabular}{lcc}
\hline \hline
Source           & 1,2 jets  & $\ge$2 jets   \\ \hline 
Data observation & 6            & 2               \\ 
SM prediction    & 6.9$\pm$1.2  & 3.3$\pm$0.5     \\
95\% C.L. Limit (events) & 6.6 & 4.4\\ \hline
&&\\
Model            &                 & \\ \hline
$W \bb$-like     & 26.4 pb         & --               \\
Top-like         & --               & 14.9 pb\\ 
\hline \hline
\end{tabular}
\end{center}
\end{table}

In summary, we observe no excess beyond the SM prediction for
heavy-flavor quark production in association with $W$ bosons in
164~$\ipb$ of data in the $\ejet$ channel and 145~$\ipb$ in the
$\mjet$ channel. Using a sample of events containing at least one jet
tagged with both the SLT and SVT algorithms, we derive 95\%
C.L. limits on anomalous heavy-flavor production
(Table~\ref{tab:clstab}). Using benchmark SM processes, we also derive
anomalous cross-section limits of 26.4~pb in a $Wb\bar{b}$-like
scenario and 14.9~pb in a top-like scenario. For comparison, the D\O\
collaboration has recently published a similar study in the form of a
search for $Wb\bar{b}$ production~\cite{wbb}. Based on the two-jet
topology, with both jets HF-tagged, that study sets a 95\% C.L. upper
cross-section limit of 6.6~pb on $W\bb$ production.

%
We thank the staffs at Fermilab and collaborating institutions, 
and acknowledge support from the 
Department of Energy and National Science Foundation (USA),  
Commissariat  \` a l'Energie Atomique and 
CNRS/Institut National de Physique Nucl\'eaire et 
de Physique des Particules (France), 
Ministry of Education and Science, Agency for Atomic 
   Energy and RF President Grants Program (Russia),
CAPES, CNPq, FAPERJ, FAPESP and FUNDUNESP (Brazil),
Departments of Atomic Energy and Science and Technology (India),
Colciencias (Colombia),
CONACyT (Mexico),
KRF (Korea),
CONICET and UBACyT (Argentina),
The Foundation for Fundamental Research on Matter (The Netherlands),
PPARC (United Kingdom),
Ministry of Education (Czech Republic),
Canada Research Chairs Program, CFI,
Natural Sciences and Engineering Research Council and 
WestGrid Project (Canada),
BMBF and DFG (Germany),
A.P.~Sloan Foundation,
Research Corporation,
Texas Advanced Research Program,
and the Alexander von Humboldt Foundation.
%

\end{document}